\documentclass[a4paper,11pt,notitlepage]{article}
\usepackage[notcite,notref]{}
\usepackage{latexsym}
\usepackage{amssymb}
\usepackage{color}
\usepackage{graphicx}
\usepackage{amsmath}
\usepackage{amsthm}
\usepackage{mathrsfs}
\usepackage{natbib}
\newtheorem{lemma}{Lemma}[section]
\newtheorem{remark}{Remark}[section]

\newtheorem{theorem}{Theorem}[section]
\newtheorem{corollary}{Corollary}[section]
\newtheorem{proposition}{Proposition}[section]

\usepackage[latin1]{inputenc}
\newcommand{\comment}[1]{}
\def\Au{\mathcal H_{1}\left(A,x^+\right)}
\def\Ar{\mathcal H_{2}\left(A\right)}
\def\UX{\mathbb{E}\left[u\left(X^{+} \right) \right]}
\def\U{U\left(X\right)}
\def\TA{\triangle\left(\mathit{A}\right)}
\def\P{\mathbb P}
\def\E{\mathbb E}
\def\Pu{\mathcal{P}_1}
\def\Pr{\mathcal{P}_2}
\def\xA{x^+\left(A\right)}

\title{Portfolio Insurance under a risk-measure constraint}
\author{Carmine De Franco\footnote{LPMA, Paris VII University, E-mail: carmine.de.franco@gmail.com}\and Peter Tankov\footnote{CMAP, Ecole Polytechnique, E-mail: peter.tankov@polytechnique.org}}
\date{}
\begin{document}
\frenchspacing

\maketitle
\begin{center}
\textbf{Abstract}
\end{center}
We study the problem of portfolio insurance from the point of view of a fund manager, who guarantees to the investor that the portfolio value at maturity will be above a fixed threshold. 
If, at maturity, the portfolio value is below the guaranteed level, a third party will refund the investor up to the guarantee. In exchange for this protection, the third party imposes a limit on the risk exposure of the fund manager, in the form of a convex monetary risk measure. 
The fund manager therefore tries to maximize the investor's utility function subject to the risk measure constraint. We give a full solution to this nonconvex optimization problem in the complete market setting and show in particular that the choice of the risk measure is crucial for the optimal portfolio to exist. Explicit results are provided for the entropic risk measure (for which the optimal portfolio always exists) and for the class of spectral risk measures (for which the optimal portfolio may fail to exist in some cases). 

\medskip


\medskip


\section{Introduction}
\numberwithin{equation}{section}
%In this paper we define a new type of portfolio insurance and look at optimal strategies for it. 
We consider the problem of a fund manager who wants to structure a portfolio insurance product where the investors pay the initial value $v_{0}$ at time $0$ and are guaranteed to receive at least the amount $z$ at maturity $T$. We assume that if, at time $T$, the value of the fund's portfolio $V_T$ is smaller than $z$, a third party pays to the investor the shortfall amount $z-V_T$. In practice, this guarantee is indeed usually provided by the bank which owns the fund. The final payoff for the investor will be 
\begin{equation}
\textrm{Payoff}=\max\left(V_T,z\right) \label {investpayoff}
\end{equation} 
In exchange, the third party imposes a limit on the risk of shortfall $-(V_T-z)^-$, represented by a law-invariant convex risk measure $\rho$. We assume that the investors' attitude to gains above the guaranteed level $z$ is modeled by a concave utility function $u$.  

The fund manager therefore faces the following problem:
\begin{align}
&\text{maximize}\quad E[u((V_T-z)^+)]\label{mp}\\
&\text{subject to}\quad R\left(V _T\right):=\rho(-(V_T-z)^-)\leq \rho_0\quad \text{and}\quad V_0 = v_0.\label{mp2}
\end{align}
The utility function applies only to the random variable $(V_T-z)^+$ as the investor is indifferent to the portfolio's value below the guarantee $z$.

This is a nonstandard maximization problem, because the objective function is not concave, and it therefore cannot be solved using standard Lagrangian methods. 
We use a technique similar to the one developed in \cite{jin.zhou.08} in the context of behavioral portfolio optimization to decouple the problem \eqref{mp}--\eqref{mp2} into two separate convex optimization problems and show that in a complete market case the optimal solution has a simple structure.

An interesting outcome of our study is that the maximization problem \eqref{mp} may not admit an optimal solution for all convex risk measures, which means that not all convex risk measures may be used to limit fund's exposure in this way. We provide conditions for the existence of the solution and show, for example, that in the Black-Scholes model, the CVaR risk measure does not satisfy these conditions.

Portfolio insurance is a widely popular concept in financial industry, and there exists an extensive literature on this topic. When the guarantee constraint is imposed in an almost sure way, a common strategy is the option based portfolio insurance, which uses put options written on the underlying risky asset as protection. The optimality of OBPI for European and American capital guarantee is studied in \cite{elkaroui.al.05}. The difficulty of finding a sufficiently long-dated option for use in OBPI has lead to the appearance of strategies which involve only the underlying risky asset, of which the most popular is the Constant Proportion Portfolio Insurance (CPPI), \citep{black.perold.92}, where the exposure to the risky asset is proportional to the difference between the value of the fund and the discounted value of the guaranteed payment. If the price path of the underlying risky asset admits jumps, the CPPI strategy no longer ensures that the fund value will be a.s. above the guaranteed level at maturity, unless the portfolio is completely deleveraged \citep{cont.tankov.09}, which usually imposes too strong a restriction on the potential gains. 

The current market practice is therefore to require that the portfolio stays above the guaranteed level with a sufficiently high probability, or, for example, that it remains above the guarantee for a certain set of stress scenarios, chosen from historical data coming from highly volatile periods. A more flexible approach, which can take into account not only the probability of loss but also the sizes of potential losses, is to impose a constraint on a risk measure of the shortfall. This has led to the development of literature on portfolio insurance and, more generally, portfolio optimization under probabilistic / risk measure constraints. 

\cite{emmer.al.01} study one-period portfolio optimization under Capital-at-Risk constraint (the Capital-at-Risk is defined as the difference between the mean value of the portfolio and its VaR). Still in the one-period setting, \cite{rockafellar.uryasev.00} provide an algorithm for minimizing the CVaR of a portfolio under a return constraint. \cite{boyle.tian.07} discuss continuous-time portfolio optimization under the constraint to outperform a given benchmark with a certain confidence level. Like us, these authors also face some issues related to the non-convexity of the optimization problem, although the non-convexity appears for a different reason (non-convexity of the constraint itself). 

Another stream of literature \citep{follmer.leukert.99,bouchard.elie.touzi.09} considers hedging problems when the hedging constraint is imposed with a certain confidence level rather than almost surely. The viscosity solution approach of \cite{bouchard.elie.touzi.09} was extended in \citep{bouchard.elie.imbert.10} to stochastic control problems under target constraint (that is, for example, under the constraint to outperform a benchmark with a certain probability) but it does not seem to be possible to treat risk measure constraints in this setting. 

\cite{he.zhou.10} have recently introduced a general methodology for solving law-invariant portfolio optimization problems by reformulating them in terms of the quantile function of the terminal value of the portfolio. While such a reformulation is in principle possible for our problem using the dual representation results for law-invariant convex risk measures (see \cite{follmer.schied.04} and \cite{jouini.al.06}), the resulting problem is still non-linear and non-convex so such a transformation does not necessarily simplify the treatment. 

\cite{gundel.weber.07} solve the problem of maximizing the (robust) utility of a portfolio under a constraint on the expected shortfall, which includes, in particular, all coherent risk measures. 
The main difference of our paper from that of Gundel and Weber, and the main novelty of our paper is that in our approach, the utility function is only applied to positive gains while the risk measure is only applied to negative shortfall. This brings us much closer to the reality of portfolio insurance and at the same time allows to obtain explicit solutions.

%Together with \cite{gundel.weber.07}, \cite{jin.zhou.08} are the closest in spirit to ours. As in Gundel and Weber we impose a kind of risk limit on the potential loss. The main difference is that they impose this risk limit on the entire portfolio measured by the utility-based shortfall whereas we apply a general convex risk measure just to the potential loss, i.e. below the guarantee. In \cite{jin.zhou.08} they do not impose any constraint in their problem, but the find out that their optimal solution is, in some way, bounded and known from the beginning in the worst scenarios. They focus their attention on the different on investor's preferences above and below gains. Their decoupling strategy is the key idea to solve our problem \eqref{mp}--\eqref{mp2}.
 
%The main novelty of our paper is that in our approach, the utility function is only applied to positive gains while the risk measure is only applied to negative shortfall. This brings us much closer to the reality and at the same time allows to obtain explicit solutions.

The rest of the paper is organized as follows. In section \ref{main} we introduce the model and optimization problem, and state the main theoretical results, including a decoupling method to solve the problem \eqref{mp} and the conditions under which this problem admits a finite solution. In sections \ref{entropic} and \ref{cvar} we investigate the case where one uses, respectively, the entropic risk measure and the spectral risk measures. The proofs of all theoretical results are postponed to section \ref{proofs}.

%%%%%%%%%%%%%%%%%%%%%%%%%%%%%%%%%%%%%%%%%%%%%%%%
\section{Main results}\label{main}
Let $\left(\Omega,\mathcal F, \mathcal F_t,\P \right)$ be a filtered probability space. 
We consider an arbitrage-free complete financial market consisting of $d$ risky assets with $(\mathcal F_t)$-adapted price processes $(S^i_t)^{i=1,\dots,d}_{0\leq t\leq T}$ and the risk-free asset with price process $S^0_t\equiv 1$. We do not specify the dynamics of risky assets and the precise definition of admissible strategies because they are not relevant for what follows. See \cite{karatzas.shreve.98} for the standard example of a market which satisfies our assumptions in the Brownian filtration. 
For an admissible trading strategy $\pi$, the investor's portfolio value is
\begin{align*}
V_T^\pi=v_0+\int_0^T \pi_u dS_u
\end{align*}
The unique martingale measure will be denoted by $\mathbb Q$, and we define $\xi:=\frac{d\mathbb Q}{d\mathbb P}$. The market completeness implies that for any $\mathcal F_T$-measurable random variable $X$ with $\E[\xi |X|]<\infty$ such that $\E\left[\xi X\right]=v_0$, there exists an admissible trading strategy $\pi$ such that $V_T^\pi:=v_0+\int_0^T\pi_t dS_t=X$ a.s.

Since the interest rate is zero, $z\leq v_0$ to avoid direct arbitrage for the investor. Moreover, without loss of generality, we will assume $z=0$ in the rest of the paper. 

The attitude of the investor towards gains above $0$ is measured, in the spirit of the Von Neumann-Morgenstern expected utility theory, by a twice differentiable, strictly concave and strictly increasing function  $u:\left[0, +\infty\right)\longrightarrow\mathbb{R}$, satisfying the usual condition $\lim_{x\to +\infty} u'(x) = 0$. We suppose $u(0)=0$ and we denote $v(y) = \sup_{x\geq 0} (u(x)-xy)$ and $I(y):=(u')^{-1}(y)$ if $y<\lim_{x\downarrow 0} u'(x)$ and $I(y)=0$ otherwise. Moreover, we assume that the following integrability condition holds: $E[ v(\lambda \xi)]<\infty$ for all $\lambda > 0$.

The risks are measured using  the convex law-invariant risk measure $\rho: \mathcal X\to \mathbb R\cup \{+\infty\}$ (see \cite{follmer.schied.04}).  The domain of definition $\mathcal X$ of $\rho$ may contain unbounded claims and may be taken equal, for example, to $L^p$ as in \cite{kaina2009convex} or a more general Orlicz space as in  \cite{biagini.frittelli.09}.
To simplify notation later on, we additionally define $\rho(X) = +\infty$ if $X\leq 0$ and $X\notin \mathcal X$. 

Using the market completeness, the optimization problem \eqref{mp}--\eqref{mp2} can be reformulated as the problem 
to find, if it exists, an $X^*\in H$ such that
\begin{equation}
\mathbb{E}\left[u\left((X^*)^{+}\right) \right]=\sup_{X\in\mathit{H}}\E\left[u\left(X^{+}\right) \right]
\label{3}
\end{equation}
where 
\begin{equation}
\label{2}
\mathit{H}:=\left\{X\in{L}^{1}\left(\xi\mathbb{P}\right)\left|\mathbb{E}\left[\xi X \right]\leq x_{0}, \rho\left(-X^{-} \right)\leq\rho_{0}\right. \right\}
\end{equation}
and $x_0=v_0$. To simplify the notation, let us define
\begin{displaymath}
\U:=\UX
\end{displaymath}
We choose $\rho_0> \rho\left(0\right)$. The problem \eqref{3} cannot be solved using classical Lagrangian methods because the function $U$ is not concave. 

Since for all $X\in\mathit{H}$,
\begin{displaymath}
\UX=\mathbb{E}\left[u\left(X\textbf{1}_{\mathit{A}}\right) \right]
\end{displaymath}
where $\mathit{A}:=\left\{X\geq 0 \right\}$, only $X\textbf{1}_{\mathit{A}}$ is important for the investor. This remark suggests the following decoupling: let  $\left(\mathit{A},x^+ \right)\in\mathcal{F}\times\mathbb{R}^+$ and consider
\begin{align}
&\Pu:\quad\textrm{maximize }\quad\quad U(Z)\quad\textrm{ subject to }Z\in\Au \label{P1}\\
\nonumber
&\Au:=\left\{Z\in{L}^{1}\left(\mathbb{\xi P}\right)\left|\right.\mathbb{E}\left[\xi Z\right]\leq x^{+},\, Z=0\, 
 \textrm{ on $\mathit{A}^{c}$, }Z\geq 0\, \textrm{ on $\mathit{A}$} \right\}\\
 \nonumber
&\textrm{and} \\
\nonumber
& \\
&\Pr:\quad\textrm{minimize }\quad\quad\mathbb{E}\left[\xi Y\right]\textrm{ subject to } Y\in\Ar \label{7}\\
\nonumber
&\Ar:=\left\{Y\in{L}^{1}\left(\mathbb{\xi P}\right)\left|\right.\rho\left(Y\right)\leq\rho_{0},\, Y=0\, \textrm{ on $\mathit{A}$, }Y\leq 0 \, \textrm{ on $\mathit{A}^{c}$}\right\}
\end{align}
For all $A\in\mathcal F$ we define:
\begin{equation}
\TA:=\inf_{Y\in\Ar}\E\left[\xi Y\right]\textrm{   and   }\xA:=x_0-\TA\label{TA}
\end{equation} 
and
\begin{equation}
U \left(A,x^+\right):=\sup_{Z\in\mathcal{H}_1\left(A,x^+\right)}U(Z) \label{U}
\end{equation}
Problem $\Pr$ is a minimization of a linear function over a convex set and, as we will see later, Problem $\Pu$ can be viewed as a concave maximization problem under a linear constraint. We will start by analysing Problems $\Pu$ and $\Pr$ and then Theorem \ref{10} will clarify the relationship between these problems and \eqref{3}.
\begin{remark}
\label{Remtrivial}
Before going on, it is important to investigate the behavior of $\Pu$ and $\Pr$ on trivial sets. If $\P\left(A\right)=0$ then $0\in \Ar$ and then $\TA\leq 0$ which means that $\xA\geq x_0\geq 0$. Therefore, $0 \in \mathcal{H}_1\left(A,\xA\right)$ and $U\left(A,\xA\right)=u(0)$.
\end{remark}

In the next lemma we will solve explicitly problem $\Pu$. 

\begin{lemma}
\label{solvP1}
Suppose $\mathbb P\left(A\right)>0$. The unique maximizer of problem $\Pu$ is given by
\begin{equation}
Z\left(A,x^+\right)=I\left(\lambda\left(A,x^+\right)\xi\right)\textbf{1}_{A}
\label{XA}
\end{equation} 
where $\lambda\left(A,x^+\right)$ is the unique solution of
\begin{equation}
\E\left[\xi I\left(\lambda\left(A,x^+\right)\xi\right)\textbf{1}_{A} \right]=x^+. \label{Lag}
\end{equation}
The value function $U(A,x^+)$ is strictly increasing and continuous in $x^+$, and for every $\lambda >0$ there exists $C<\infty$ such that 
\begin{align}
U(A,x^+) \leq C+ \lambda x^+\label{boundU}
\end{align}
for all $A \in \mathcal F$ and all $x^+\geq 0$.
\end{lemma}
The next example will clarify the role of $\TA$. Fix $A$ such that $\P\left(A\right)>0$ and suppose $\TA=-\infty$. It is then possible to find,  for each $n\in\mathbb{N}$ a random variable $Y^{n}\in\Ar$ such that $\E\left[\xi Y^{n} \right]\leq-n$. Define now
\begin{displaymath}
X^n=\frac{x_{0}+n}{\E\left[\xi \textbf{1}_{\mathit{A}} \right]}\textbf{1}_{\mathit{A}} + Y^n
\end{displaymath}
It is clear that $X^n\in\mathit{H}$ for all $n$ and $U\left(X^n\right)\rightarrow \sup_x u\left(x\right)$, which means that Problem \eqref{3} does not admit a maximizer.
%From a financial point of view, the case $\TA=-\infty$ is singular but, nevertheless, interesting: first remark that $Y^n\leq 0$ a.s. In this case the fund manager can sell the contract $Y^n$ for at least the amount $n$ keeping his risk exposure bounded by $\rho_0$, and use this amount to overperform the Investor's wealth utility. Following this strategy, the fund manager will infinitely increase the Investor's \emph{happiness}, but may lead to the ruin of third part (generally the Insurance company) on the scenarios $\omega\in A^c$, because it has to pay the amount $Y^n\left(\omega\right)$ which can be really important when $n\rightarrow +\infty$. This singular case is due to the wrong choice of $\rho$ as the measure of risk for the Insurance company. 
To avoid this problem, we shall use one of the following assumptions on $\triangle$:
\begin{align}
&\text{for all }A \in \mathcal F,\, \TA>-\infty 
\label{ass1}\\
&\inf_{A\in \mathcal F} \TA>-\infty.\label{ass1a}
\end{align}
Clearly, \eqref{ass1} depends on the particular choice of $\rho$ and $\xi$. In particular, a choice under which $\TA=-\infty$ for some $A$ is not appropriate in this kind of portfolio insurance. As we will see later in the example we will present, for the $\textrm{CVaR}_{\lambda}$ risk measure in the Black and Scholes model, $\TA=-\infty$ whereas the same risk measure coupled with a bounded $\xi$ satisfies \eqref{ass1a}.

Assumptions \eqref{ass1} and \eqref{ass1a} can be difficult to check; the following condition, which is simpler, guarantees \eqref{ass1a} but it is not necessary.
\begin{proposition}
\label{66}
Condition \eqref{ass1a} is implied by the condition
\begin{equation}
\gamma_{min}\left(\xi\mathbb{P}\right)<+\infty,
\label{68}
\end{equation}
where $\gamma_{min}$ is the minimal penalty function of $\rho$ defined by
$$
\gamma_{min}(\mathbb Q) = \sup_{X\in \mathcal A_{\rho}} \E^{\mathbb Q}[-X],
$$ 
where $\mathcal A$ is the acceptance set of $\rho$. 
\end{proposition}
%As we said, this condition is not necessary: we will see an example where, under some conditions, we may have 
%$\gamma_{min}\left(\xi\mathbb{P} \right)=+\infty$ even if problem \eqref{3} has a finite solution.

The following result clarifies the relationship between Problem \eqref{3} and $\Pu$--$\Pr$, giving us a method to solve the former.
\begin{theorem}
\label{10}
Let \eqref{ass1} hold. 
Then,
\begin{equation}
\sup_{X\in H}\U=\sup_{A\in\mathcal{F}}U\left(\mathit{A},\xA \right).
\label{attainsup}
\end{equation}
If, in addition, \eqref{ass1a} holds, then both sides of \eqref{attainsup} are finite.
\end{theorem}

Theorem \eqref{10} gives us a condition under which the value function of problem \eqref{3} is finite and a way to compute it: \\
\textbf{Algorithm 1 }:
\begin{enumerate}
  \item fix $\mathit{A}\in\mathcal{F}$
  \item solve $\mathcal{P}_{2}\left(\mathit{A} \right)$ and find $\triangle\left(\mathit{A} \right)$
  \item solve $\mathcal{P}_{1}\left(\mathit{A} \right)$ with parameter $\left(A,\xA=x_{0}-\TA\right)$
  \item maximize the value function of $\Pu$, $U\left(\mathit{A},\xA \right)$ (repeating the steps 1--3), over $A\in\mathcal F$
\end{enumerate}

The next result establishes a link between the maximizers of problem \eqref{3} and $\Pu$--$\Pr$. 
\begin{theorem}
\label{corOpt}
Let \eqref{ass1} hold. 

Suppose that $X^*$ achieves the maximum in Problem \eqref{3} and define $A^* :=\left\{X^*\geq 0\right\}$. One has
\begin{itemize}
\item  $A^*$  achieves the maximum in the right-hand side of \eqref{attainsup}
\item $Y^*:=X^* - X^*\textbf{1}_{A^*}\in\mathcal{H}_{2}\left(A^*\right)$ achieves the minimum in $\Pr$.
\end{itemize}

Conversely, let $A^*\in\mathcal F$, $\P\left(A^*\right)>0$ and $Y^*\in\mathcal H_2\left(A^*\right)$ such that 
\begin{align}
\nonumber
&U\left(A^*,x^+\left(A^*\right)\right)=\sup_{A\in\mathcal F}U\left(\mathit{A},\xA  \right)\\
\nonumber
&\E\left[\xi Y^* \right]=\triangle\left(A^*\right)=\inf_{Y\in\mathcal{H}_2\left(A^*\right)} \E\left[\xi Y\right]
\end{align}
Then a solution of problem \eqref{3} is given by
\begin{equation}
\label{VF}
X^*:= I\left(\lambda^*\xi \right)\textbf{1}_{A^*}+ Y^*
\end{equation}
where $\lambda^*=\lambda\left(A^*,x^+\left(A^*\right)\right)$ verifies \eqref{Lag}. 
In this case, the payoff for the investor will be
\begin{equation}
\label{payoffopt}
\textrm{Payoff}=I\left(\lambda^*\xi \right)\textbf{1}_{A^*}
\end{equation}
\end{theorem}

\begin{remark}\label{nomin}
Algorithm 1 and Theorem \ref{corOpt} give us a way to find an optimal solution for problem \eqref{3} if we are able to find a maximizer in \eqref{attainsup} and the minimizer in $\Pr$.

But what happens in the case when the maximizer in \eqref{attainsup} or the minimizer in $\Pr$ does not exist? In this case, under Assumption \ref{ass1}, following the steps of the proof of Theorem \ref{attainsup}, one can show that for all $\varepsilon>0$ there exist $A^\varepsilon\in\mathcal F$, $\lambda^\varepsilon\in\mathbb R$ and $Y^\varepsilon\in\mathcal{H}_2(A^\varepsilon)$ such that
\begin{equation}
\label{Xesp}
X^\varepsilon:= \left[I\left(\lambda^\varepsilon \xi\right)\right]\textbf{1}_{A^\varepsilon} + Y^\varepsilon
\end{equation}
verifies $U\left(X^\varepsilon\right)+\varepsilon>\sup_{X\in\mathit{H}}\U$.
\end{remark}

The main difficulty in applying Theorems \ref{10} and \ref{corOpt} is how to find a maximizer $A^*$. Generally, maximization over the sets in $\mathcal F$ is not simple. Our aim here is to show that this latter maximization may be carried out over a subset of $\mathcal F$, parameterized by a real number. A similar approach was taken in \cite{jin.zhou.08}.

We already know, from Theorem \ref{10}, that
\begin{displaymath}
\sup_{X\in H}\U =\sup_{A\in\mathcal F}U\left(A,\xA\right)=\sup_{A\in\mathcal F} \sup_{X\in\mathcal H_1\left(A,\xA\right)}\U
\end{displaymath}
In order to focus our attention on the set dependence, we will introduce the following notation:
\begin{equation}
\label{VA}
v\left(A\right):=\sup_{X\in\mathcal H_1\left(A,\xA\right)}\U 
\end{equation}
Let us also define
$\underline{\xi}:=essinf\, \xi$ and $\overline{\xi}:=esssup\, \xi$. 
\begin{theorem}
\label{15}
Suppose that the law of $\xi$ has no atom and let $A\in\mathcal F$. Let $c\in\left[ \underline{\xi},\overline{\xi} \right]$ such that $\P\left(\xi\leq c\right)=\P\left(A\right)$. Then  
\begin{equation}
v\left(A\right)\leq v\left(\left\{\xi\leq c \right\} \right)
\label{maxinc}
\end{equation}
which means that
\begin{equation}
\sup_{X\in H}\U=\sup_{A\in\mathcal{F}}v\left(\mathit{A}\right)=\sup_{c\in\left[\underline{\xi},\overline{\xi} \right]}v\left(\left\{\xi\leq c \right\}\right). 
\label{16}
\end{equation}
\end{theorem}

In order to make the notation simpler, let $v\left(c\right):=v\left(\left\{\xi\leq c\right\}\right)$.
With this result, we can make Algorithm 1 simpler:\\
\textbf{Algorithm 2}:
\begin{enumerate}
  \item fix $c\in\left[\underline{\xi},\overline{\xi} \right]$ and consider $\mathit{A}=\left\{\xi\leq c\right\}$\\
  \item solve $\Pr$ with parameter $\left(\left\{\xi\leq c\right\}\right)$ and find $\triangle\left(c \right):=\triangle\left(\left\{\xi\leq c\right\}\right)$\\
  \item solve $\Pu$ with parameters $\left(\left\{\xi\leq c\right\},x_{0}-\triangle\left(c \right)\right)$\\
  \item find $c^{*}$ that maximizes $c\mapsto v\left(c\right)$\\
\end{enumerate}
%The following results shows how, under mild conditions, this maximum exists.
%\begin{proposition}${}$
%\label{continuity_v} 
%Suppose that the function $c\rightarrow \triangle\left(c\right)$ is continuous on $\left[\underline\xi,\, \overline\xi\right]$. Then $v$ is also continuous on the same set and
%there exists $c^*\in\left(\underline\xi,\, \overline\xi\right)$ which achieves the supremum in \eqref{16}
%\end{proposition}
%Under the hypothesis of Proposition \ref{continuity_v}  we can use \textbf{Algorithm 2} in order to solve Problem \eqref{3}. Moreover, if conditions in Theorem  \ref{corOpt} hold then there exist a maximizer in Problem \eqref{3}. Later on, we shall give examples in which $c\rightarrow \triangle\left(c\right)$ is continuous.

The question of the existence of $c^*$ which maximizes $c\mapsto v(c)$, and the related question of the existence of the optimal pay-off for the fund manager is difficult to answer for general risk measures. A complete answer to this question will be given in section \ref{entropic} in the case of the entropic risk measure (see Theorem \ref{entropic.exist}) and in section \ref{cvar} for spectral risk measures (Theorem \ref{cond_spect}). 
%%%%%%%%%%%%%%%%%%%%%%%%%%%%%%%%%%%%%%%%%%%%%%%
\section{Example: entropic risk measure}
\label{entropic}
%%%%%%%%%%%%%%%%%%%%%%%%%%%%%%%%%%%%%%%%%%%%%%%

In this section we show how Theorems \ref{corOpt} and \ref{15}  can be used to solve problem \eqref{3} when the risk measure in question is the entropic risk measure (ERM) defined by
\begin{equation}
\rho_{\beta}\left(X\right):=\beta\ln\mathbb{E}\left[\exp\left(-\frac{1}{\beta}X \right) \right]
\label{36}
\end{equation}
where $\beta>0$. Throughout this and the following section, we use the notation of Section \ref{main} and suppose that all the assumptions stated in the beginning of that section stand in force. 

As shown in Example 4.33 in \cite{follmer.schied.04} (see also section 5.4 in \cite{biagini.frittelli.09} for the case of unbounded claims), the entropic risk measure can be represented as
\begin{displaymath}
\rho_{\beta}\left(X\right)=\sup_{\mathbb Q\ll\P,\, \log\left(\frac{d \mathbb Q}{d\P}\right)\in L^1\left(\mathbb Q\right)}\left(\E_{\mathbb Q}\left[-X\right]- \beta\E_{\mathbb{Q}}\left[\log\left(\frac{d\mathbb{Q}}{d\P} \right) \right]\right).
\end{displaymath}
In particular, $\gamma_{min}\left(\xi\P\right)=\beta\E\left[\xi\log\left(\xi\right)\right]$. %Throughout this section we assume that $\xi\log\left(\xi\right)\in L^1\left(\P\right)$, which, by Proposition \ref{66}, guarantees that the condition \eqref{ass1a} is satisfied.
\begin{theorem}\label{entropic.exist}
Let the risk measure $\rho$ be given by \eqref{36} and assume that the state price density $\xi$ has no atom and satisfies $\xi\log\xi\in\mathbb{L}^{1}\left(\mathbb{P}\right)$. Then the optimal payoff for the fund manager is given by 
\begin{equation}
	V^{*}:=I\left(\lambda\left(c^{*}\right)\xi\right) \textbf{1}_{\left\{\xi\leq c^{*}\right\}}
	-\beta\left[\log\left(\frac{\beta}{\eta\left(c^*\right)}\xi\right)\right]^+ \textbf{1}_{ \left\{\xi>c^*\right\}}
	\nonumber
\end{equation}
where
	\begin{itemize}
	 \item $\lambda\left(c\right)$ is the unique solution of $\E\left[\xi I\left(\lambda\left(c\right)\xi\right) \textbf{1}_{\left\{\xi\leq c\right\}} \right]=x_{0}-\triangle\left(c\right)$
	\item $\alpha\left(c\right)=\P\left(\xi>c\right)$
 \item $\triangle\left(c\right)= -\beta \mathbb E\left[\xi \log\left(\frac{\beta \xi }{\eta(c)}\vee 1\right)\right]$
	\item $\eta\left(c\right)$ is the unique solution of:
$\mathbb E\left[\left(\frac{\beta \xi}{\eta(c)}\vee 1\right) \mathbf 1_{\xi > c}\right] = e^{\frac{\rho_0}{\beta}} + \alpha(c)-1$.
	\item $c^{*}$ attains the supremum of $c\rightarrow\E\left[u\left(I\left(\lambda\left(c\right)\xi\right)\right)\textbf{1}_{\left\{\xi\leq c\right\}} \right]$
  \end{itemize}
\label{69}
\end{theorem}
\paragraph{Numerical example}

We will apply Theorem \ref{69} in a simple case. Let the market be composed of one risky asset, $S$, which follows the Black and Scholes dynamics:
\begin{displaymath}
dS_t=S_t\left( bdt+\sigma dW_t\right)\, \quad S_0>0
\end{displaymath}
Suppose $\mu=b/\sigma>0$. The unique equivalent martingale measure is given by $\mathbb Q=\xi \P$, where $\xi=\exp(-\mu W_T -\mu^2 T/2)=\left[S_T \exp\left(T\left(\sigma^2-b\right)/2\right)/S_0 \right]^{-\frac{b}{\sigma^2}}$.\\
We will use the exponential utility function $u\left(x\right)=1-e^{-\delta x}$. For this example we take $b=0.15$, $\sigma=0.4$, $\mu=0.375$, $T=1$, $S_0=5$, $x_0=1.5$, $\rho_0=1.5$, $\beta=1$, and $\delta=0.6$.

\begin{figure}
\centerline{\includegraphics[width=1\textwidth]{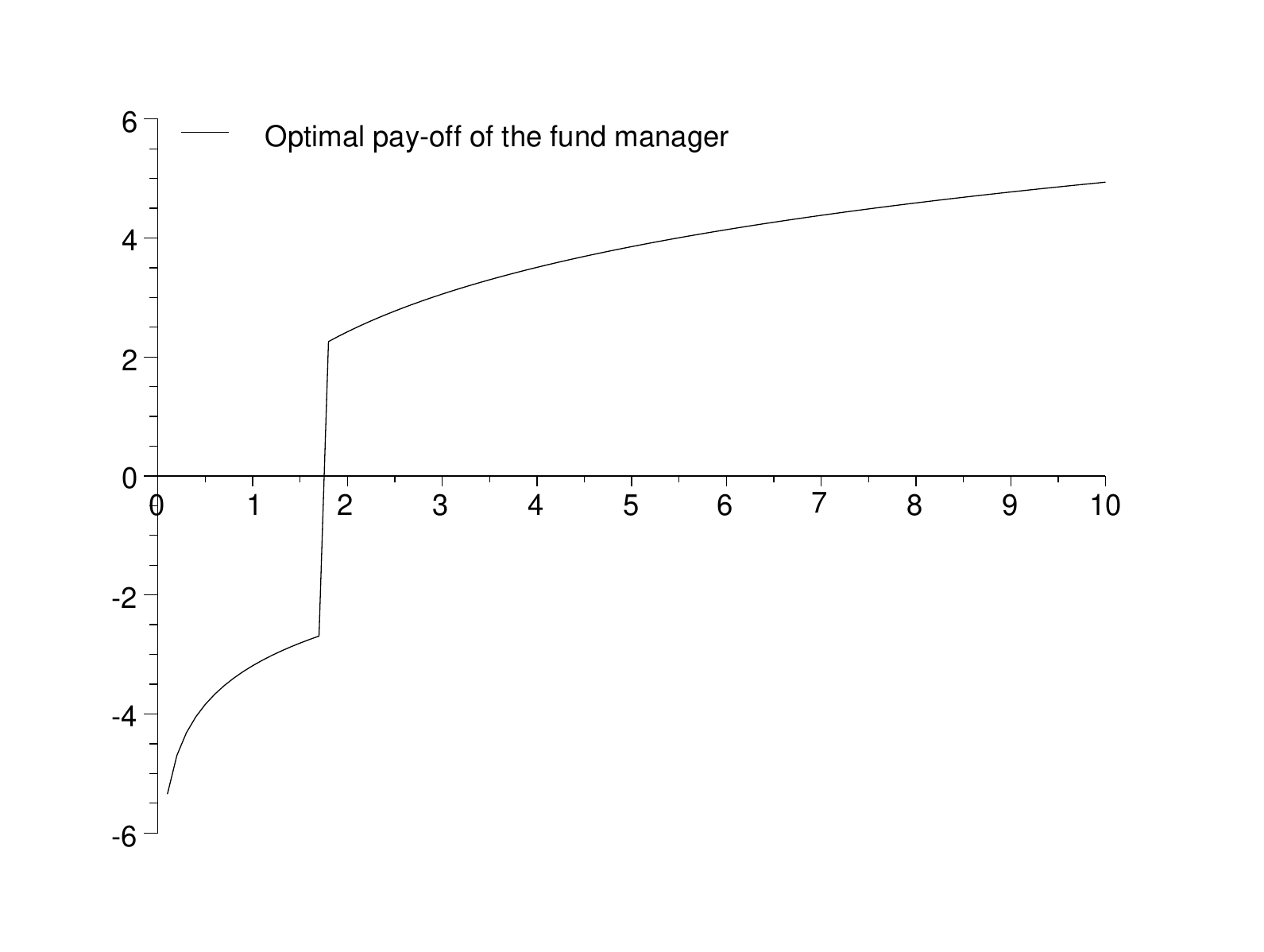}}
\caption{Optimal pay-off of the fund manager as function of the stock price value $S_T$.}
\label{optpayoff.fig}
\end{figure}

\begin{figure}
\centerline{\includegraphics[width=0.9\textwidth]{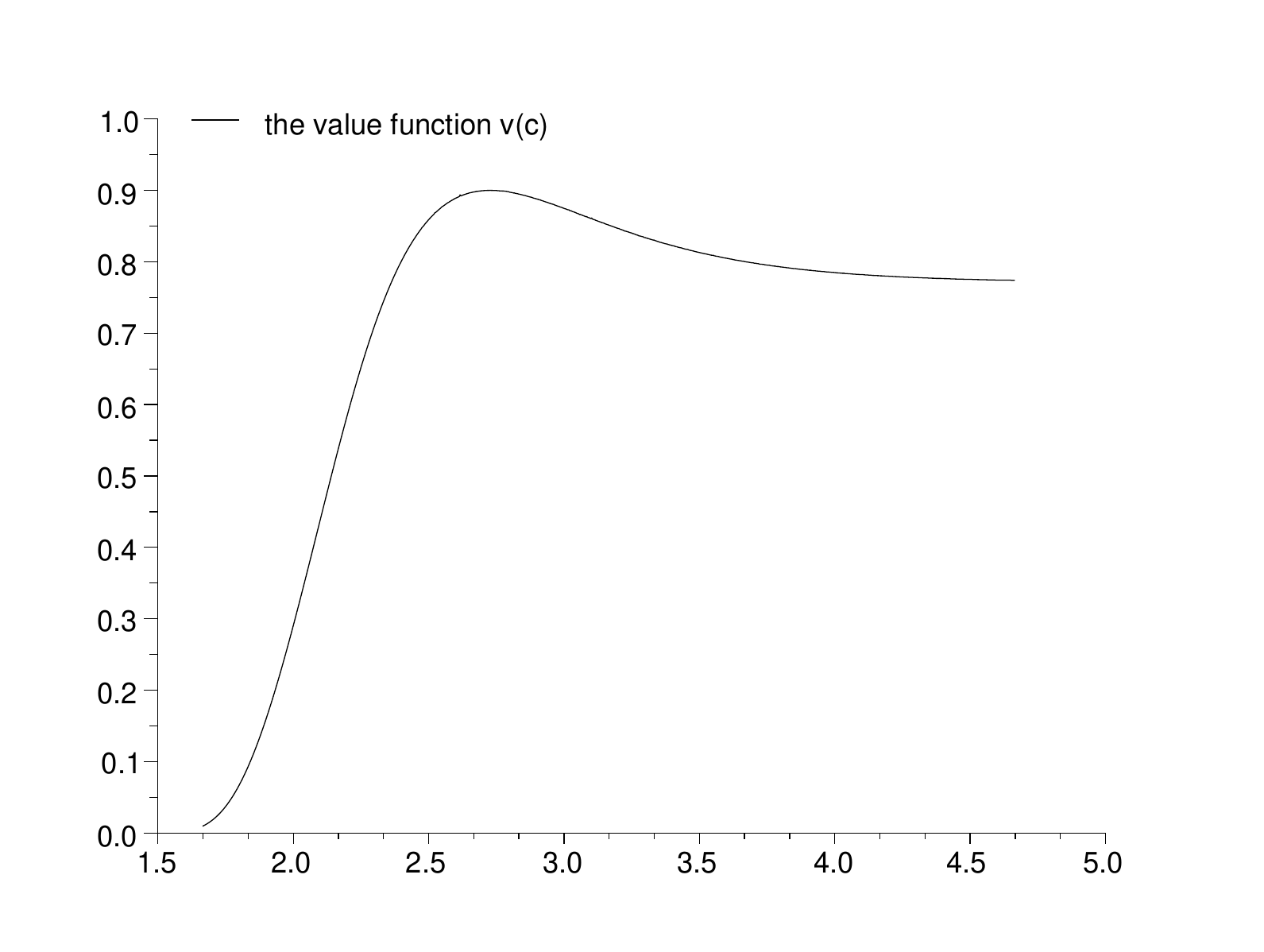}}
\caption{Value function of Problem $\mathcal P_1$ as function of $c$.  }
\label{valfunc}
\end{figure}

\begin{figure}
\centerline{\includegraphics[width=0.9\textwidth]{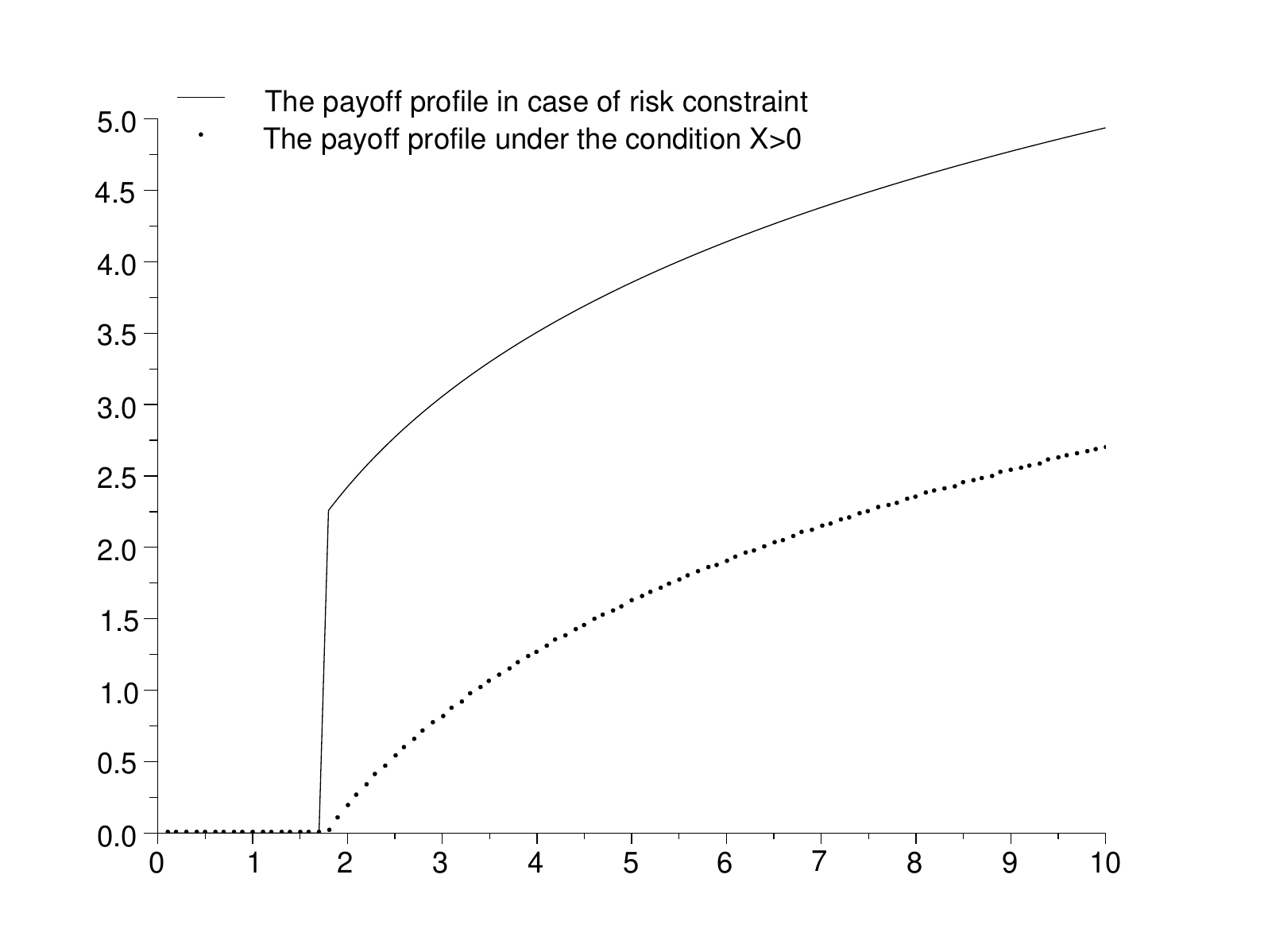}}
\caption{The gain obtained by allowing a risk tolerance. The solid curve shows the optimal pay-off for the investor as in \eqref{investpayoff} and the dotted curve the optimal one when no risk is allowed: $\max\E\left[1-e^{-\delta X^+}\right]$ under $\E\left[\xi X \right]=x_0$ and $X\geq 0$.
}
\label{invpayoff.fig}
\end{figure}

The optimal pay-off is a spread of two options on the log contract $\log (S_T)$: one option is sold to match the desired risk tolerance and the second one is bought to obtain the gain profile desired by the investor. 
\begin{equation} 
X^*:=\left[\frac{L}{\delta}\log\left(S_T\right) + K_1 \right]^+ \textbf{1}_{\left\{S_T\geq s^*\right\}} -\beta\left[K_2-L\log\left(S_T\right) \right]^+ \textbf{1}_{\left\{S_T<s^*\right\}}
\end{equation}
where the numerical values of the constants are
\begin{align}
\nonumber
& s^*=S_0 \exp\left(T\left(b-\sigma^2\right)/2\right) \left(c^*\right)^{-\frac{-\sigma^2}{b}}=1.70907\\
\nonumber
& L=\frac{b}{\sigma^2}=0.9375\\
\nonumber
& K_1=\frac{1}{\delta}\left(\frac{b\left(\sigma^2-b\right)}{2\sigma^2}T-\frac{b}{\sigma^2}\log\left(S_0\right)-\log\left(\frac{\lambda\left(c^*\right)}{\delta}\right) \right) = 1.34026\\
\nonumber
& K_2=\frac{b}{\sigma^2}\log\left(S_0\right)-\frac{b\left(\sigma^2-b\right)}{2\sigma^2}T +\log\left(\frac{\beta}{\eta\left(c^*\right)} \right)=3.18886\\
\nonumber
& c^*= 2.72293 \\
\nonumber
& \lambda\left(c^*\right)= 0.0596571\\
\nonumber
& \eta\left(c^*\right)= 0.185501\\
\nonumber
\end{align}
The optimal pay-off of the fund manager as function of $S_T$ is shown in Figure \ref{optpayoff.fig}. Figure \ref{invpayoff.fig} shows the gain for the investor compared to the situation where no risk is allowed. 
The (opposite of) extra capital made available due to the risk tolerance is given by $\triangle\left(c^*\right)=-1.17387$ and the probability of no loss is $\P\left(S_T\geq s^*\right)=0.946722$. Finally, the optimal value function for $\mathcal P_1$ is $ v\left(c^*\right)=0.900134$. Figure \ref{valfunc} shows the value function as function of $c$. 

%%%%%%%%%%%%%%%%%%%%%%%%%%%%%%%%%%%%%%%%%%%%%
\section{Example: $CVaR$ and spectral risk measures}
\label{cvar}
%%%%%%%%%%%%%%%%%%%%%%%%%%%%%%%%%%%%%%%%%%%%%

%In our second example we first use a spectral risk measure in Problem \eqref{3}, in order to show a condition under which Assumption \ref{assTA} hold and then we apply it in the special case of the $CVaR$. 
The $CVaR_{\beta}$ is a coherent risk measure defined by
\begin{equation}
CVaR_{\beta}\left(X \right):=\frac{1}{\beta}\int_{0}^{\beta}VaR_{u}\left(X\right)du = -\frac{1}{\beta} \int_0^\beta F^{-1}_X (u) du,
\label{CVaR}
\end{equation}
where $F^{-1}_X$ is a generalized inverse distribution function of $X$. Since the generalized inverse distribution function has at most a countable number of discontinuities, this definition does not depend on the particular choice of this function (right-continuous or left-continuous). In this section we shall always use the definition
\begin{align}
F_X^{-1}(u) := \inf\{x: F(x)\geq u\}\label{geninv.eq}
\end{align}
with the convention $\inf \emptyset = +\infty$. 

The $CVaR$ is the building block for a wide class of coherent risk measures called \emph{spectral risk measures}. Given a probability measure $\mu$ on $\left[0,1\right]$, the spectral risk measure $\rho_\mu$ is defined by
\begin{align}
&\rho_\mu\left(X \right):=\int_{0}^{1}CVaR_{u}\left(X\right)\mu\left(du\right) = \int_0^1 \phi\left(u\right)VaR_{u}\left(X\right) du
\label{rho_mu}
\end{align}
where 
\begin{align}
\phi\left(x\right):=\int_x^1 \frac{\mu\left(ds\right)}{s}
\label{phi_mu}
\end{align}
The function $\phi$ is right-continuous, nonincreasing and by Fubini's Theorem, $\int_0^1\phi\left(x\right)dx=1$. The case $\mu\left(du\right)=\delta_\beta\left(du\right)$ corresponds to $CVaR_\beta$. The function $\phi$ completely characterizes the spectral risk measure $\rho_\mu$. 

In this section, we solve the portfolio optimization problem when the risk constraint is given by a spectral risk measure. We first need to compute the mappings $A\to\TA$ and $c\mapsto \Delta(c)$.
\begin{lemma}
\label{explicit_TA}
For $A\in\mathcal F$ with $\P\left(A\right)<1$, let $\hat F_\xi$ be the conditional distribution of $\xi$ on $A^c$ and define $\alpha_A:=\P\left(A^c\right)$. 
$\TA<-\infty$ if and only if 
\begin{align}
&\lim_{x\rightarrow 0^+}\frac{\hat F_\xi ^{-1}\left(1-x\right)}{\phi\left(x\right)}<+\infty
\label{cond_spect_lim}
\end{align}
In this case
\begin{eqnarray}
\TA&=& - \rho_0 \max_{x\in\left[0,1\right]} r\left(x\right)\label{TA_value}\\
r\left(x\right)&:=&\frac{\alpha_A}{\int_0^{\alpha_A x}\phi\left(u\right)du}\int_0^x \hat F_\xi^{-1}\left(1-u\right)du \label{rx}\\
\nonumber
\end{eqnarray}
\end{lemma}
\begin{corollary}
\label{max_c_spectr}
The function $\Delta (c)$ is given by
$$
\Delta(c) = -\rho_0 \max_{0\leq z \leq \alpha(c)} R(z),\quad R(z):=\frac{\E\left[\xi \textbf{1}_{\left\{ 1-F_\xi(\xi)<z \right\}} \right]}{\int_0^{z}\phi\left(u\right)du}
$$
Assume that the limit
\begin{align}
\lim_{x\rightarrow 0^+}\frac{F_\xi ^{-1}\left(1-x\right)}{\phi\left(x\right)}
\label{condlim}
\end{align}
exists. Then
$$
\lim_{c\uparrow \overline \xi} \Delta(c) = -\rho_0 \frac{F_\xi ^{-1}\left(1-x\right)}{\phi\left(x\right)}.
$$
\end{corollary}
\noindent The following theorem, which is the main result of this section, characterizes the solution of the problem \eqref{3} when the risk constraint is given by a spectral risk measure via a one-dimensional optimization problem. 
\begin{theorem}
\label{cond_spect}
Assume that there exists $c^*$ with $\P[\xi>c^*]>0$ such that $v(c^*) = \max_{\underline\xi \leq c \leq \overline \xi} v(c)$ with 
\begin{align*}
v(c) = \E[u(I(\lambda(c)\xi))\mathbf{1}_{\xi \leq c}],
\end{align*}
where $\lambda(c)$ is the solution of 
$$
\E[\xi I(\lambda(c)\xi) \mathbf{1}_{\xi \leq c} ] = x_0 + \frac{\rho_0\E[\xi \mathbf 1_{\xi>c}]}{\int_0^{\P[\xi>c]} \phi(u)du}.
$$
Then the solution to the problem \eqref{3} is given by 
$$
X^* = I(\lambda(c^*)\xi)\mathbf 1_{\xi\leq c^*} - \frac{\rho_0}{\int_0^{\P[\xi > c^*]} \phi(u)du}\mathbf 1_{\xi>c^*}.
$$
\end{theorem}
\begin{remark}
\label{remark_limit_x0}
If $\sup_{\underline \xi \leq c \leq \overline \xi} v(c)$ is attained only by $c^* = \overline \xi$ and 
$$
\lim_{c\to \overline \xi} \frac{\E[\xi \mathbf 1_{\xi>c}]}{\int_0^{\P[\xi>c]} \phi(u)du} <\infty,
$$
(the latter condition holds, in particular, if $\overline \xi <\infty$), 
then $\inf_{A\in \mathcal F} \Delta(A) >-\infty$ but this infimum is not achieved: the extra gain from allowing a risk tolerance is bounded, but the optimal claim does not exist. Intuitively, claims which are ``almost optimal'' will lead to a very large loss occurring with a very small probability. 

If 
$$
\limsup_{c\to \overline \xi} \frac{\E[\xi \mathbf 1_{\xi>c}]}{\int_0^{\P[\xi>c]} \phi(u)du} =\infty
$$
then $\inf_{A\in \mathcal F} \Delta(A) =-\infty$: the extra gain from allowing a risk tolerance is unbounded.
\end{remark}
\paragraph{The special case: $CVaR_\beta$}${}$\\
In this case,  $\mu\left(du\right)=\delta_\beta\left(du\right)$ and $\phi_\beta\left(x\right):=\frac{1}{\beta}\textbf{1}_{\left\{\beta>x\right\}}$, and the limit appearing in Condition \eqref{cond_spect_lim} in Lemma \ref{explicit_TA} becomes
$$
\lim_{x\to 0^+}  \beta\hat F_{\xi}^{-1}(1-x) =  \beta\overline{\xi}
$$
Corollary \ref{max_c_spectr}, Lemma \ref{solvP1} and Theorems \ref{15} and \ref{cond_spect} enable us to give the solution of Problem \eqref{3}: 
\begin{itemize}
\item If $\bar \xi := \mathrm{essup}\, \xi(\omega) < \infty$, then the value function of problem \eqref{3} is:
\begin{equation}
\label{valuefunCVaR}
\sup_{X\in H}\U=\sup_{c\in\left[\underline{\xi},\, \overline{\xi} \right]}\E\left[u\left(I\left(\lambda\left(c\right)\xi\right)\right)\mathbf{1}_{\left\{\xi\leq c\right\}} \right]
\end{equation}	
where $\lambda\left(c\right)$ is the unique solution of 
$$
\E\left[\xi I\left(\lambda\left(c\right)\xi\right)\mathbf{1}_{\left\{\xi\leq c\right\}} \right]=x_{0}+\rho_{0} \frac{\mathbb E[\xi \mathbf 1_{\{\xi > c\}}]}{1\wedge \frac{\alpha(c)}{\beta}}
$$
\item If $\bar \xi = +\infty$ then there exists $A\in \mathcal F$ with $\TA=-\infty$.
\end{itemize}
The maximum in \eqref{valuefunCVaR} is always attained for some $c^*\in[\underline\xi,\, \overline{\xi}]$ because the value function is continuous and $[\underline\xi,\overline\xi]$ is compact. If $c^*<\overline\xi$ then Theorem \ref{cond_spect} applies and then we have a optimal solution for Problem \eqref{3}. If the maximum is attained at $c^*=\overline\xi$, then, as in Remark \ref{remark_limit_x0}, the optimal claim does not exist.
\begin{remark}
From Theorem $4.47$ in \cite{follmer.schied.04}, the minimal penalty function for the $CVAR_{\beta}$ is given by:
\begin{displaymath}
\gamma_{min}\left(\mathbb{Q}\right):=\left\{ \begin{array}{ll}
0 & \textrm{if } \frac{d\mathbb{Q}}{d\P}\leq \frac{1}{\beta},\quad \P\textrm{-a.s}\\
+\infty& \textrm{otherwise}
\end{array}\right.
\end{displaymath}
If $\xi$ is bounded but $\P\left(\xi>\frac{1}{\beta}\right)>0$ then $\gamma_{min}\left(\xi\mathbb{P} \right)=+\infty$  and we have an example of a situation where Assumption \eqref{ass1a} holds but the stronger assumption \eqref{68} fails. 
\end{remark}

%%%%%%%%%%%%%%%%%%%%%%%%%%%%%%%%%%%%%%%%%%%%%
\section{Proofs}
\label{proofs}
%%%%%%%%%%%%%%%%%%%%%%%%%%%%%%%%%%%%%%%%%%%%%

\subsection{Proof of Lemma \ref{solvP1}}
Introduce the new probability space $\left( A, \mathcal F_A:=\{B\cap A, B\in \mathcal F\},\mathbb P\left(\cdot\left|\right. A\right)\right)$ and let $\E_{A}$ denote the expectation under the conditional probability $\P\left(\cdot\left|\right. A\right)$. The maximizer of $\Pu$, if it exists, is given by
\begin{displaymath}
Z\left(A,x^+\right)=W\left(A,x^+\right) \textbf{1}_{A}
\end{displaymath}
where $W(A,x^+)$ is the maximizer of the following problem on the new probability space: 
\begin{align}
\nonumber
&\sup_{W\geq 0}\E_A\left[ u\left(W\right)\right]\\
\nonumber
&\textrm{subject to }\E_A\left[\xi W \right]=\frac{x^+}{\P\left(A\right)}.%:=\delta\left(A,x^+\right)\\
\nonumber
\end{align}
This is a classical problem of maximizing a concave function under a linear constraint which can be solved by Lagrangian methods (see e.g., \cite{karatzas.shreve.98}). First, $v$ is continuously differentiable, and since the mapping $\lambda\mapsto \E[v(\lambda\xi)]$ is convex and finite for all $\lambda$, it is differentiable, and using Fatou's lemma we get that 
 $\E[\xi v'(\lambda \xi)] = \E\left[\xi I\left(\lambda \xi\right)\right]<+\infty$ for all $\lambda>0$. Therefore, the solution to the above optimization problem is
\begin{align*}
&W\left(A,x^{+}\right)= I\left(\lambda\left(A,x^{+}\right)\xi \right)
\end{align*}
where $\lambda\left(A,x^{+}\right)$ is the unique solution of $\E_A\left[\xi I\left(\lambda \xi\right)\right]=\frac{x^+}{\P\left(A\right)}$.

To show that $x^+\mapsto U\left(A,x^+\right)$ is strictly increasing,  let $x_1^+<x^+_2$.  Then the random variable 
$$
X=I\left(\lambda\left(A,x_1^+\right)\xi\right)\textbf{1}_A +\frac{x^+_2-x_1^+}{\E\left[\xi \textbf{1}_A\right]}
$$
belongs to $\mathcal H_1\left(A,x_2^+\right) $, which proves that $U\left(A,x^+_1 \right) < U\left(A,x^+_2 \right)$. 

The continuity of $U$ follows from inequality
\begin{align}
u(I(\lambda \xi)) \leq v(\xi) + \xi I(\lambda \xi) \label{fenchel.eq}
\end{align}
and the continuity of $x^+\mapsto \lambda(A,x^+)$, which is straightforward since the function $\lambda\mapsto \E[\xi I(\lambda\xi)\mathbf 1_{A}]$ is strictly decreasing and continuous. 
The upper bound on $U$ is also a consequence of \eqref{fenchel.eq}, after taking expectations. 
\subsection{Proof of Proposition \ref{66}}
By definition of $\gamma_{min}$, 
\begin{align*}
\gamma_{min}\left(\xi\P\right)& = \sup_{Y\in\mathcal{A}_{\rho}}\E\left[-\xi Y \right]\\
&=\sup_{Y+\rho_{0}\in\mathcal{A}_{\rho}}\E\left[-\xi Y \right]-\rho_{0}\\
& \geq  \sup_{Y+\rho_{0}\in\mathcal{A}_{\rho},Y\leq 0}\E\left[-\xi Y \right]-\rho_{0} \\
& \geq  \sup_{Y+\rho_{0}\in\mathcal{A}_{\rho},\ Y\leq 0,\  Y=0\ \text{on}\ A}\E\left[-\xi Y \right]-\rho_{0}\\&
 = \sup_{Y \in \mathcal H_2(A)}\E\left[-\xi Y \right]-\rho_{0} = -\TA - \rho_0,
\end{align*}
from which the result follows. 

\subsection{Proof of Theorem \ref{10}}
Let us first prove \eqref{attainsup}. We start with the inequality ``$\leq$''. 
Let $X^n\in H$ such that $U\left(X^n\right)\uparrow\sup_{X\in H}\U$. Define $A_n:=\left\{X^n\geq 0\right\}$ and $x_n:=\mathbb E \left[\xi X^n \textbf{1}_{A_n}\right]$.\\
One has then
\begin{eqnarray}
\nonumber
U\left(X^n\right)&=&U\left(X^n\textbf{1}_{\mathit{A}^n}\right)\\
\nonumber
&\leq & U\left(\mathit{A}^n,x_n \right)\\
\nonumber
& \leq & U\left(\mathit{A}^n,x_{+}\left(\mathit{A}_n  \right) \right)\\
\nonumber
& \leq &  \sup_{A\in\mathcal{F}}U\left(\mathit{A},x_{+}\left(\mathit{A}\right) \right),
\nonumber
\end{eqnarray}
The first inequality holds because $X^n\textbf{1}_{\mathit{A}^n}\in\mathcal{H}_1 \left(A_n,x_n\right)$  and $U(A^n,x_n)$ is the sup over $\mathcal{H}_1 \left(A_n,x_n\right)$. The second inequality follows from the fact that $U(A,x^+)$ is nondecreasing in $x^+$ provided we can prove that $x_n \leq x_{+}\left(\mathit{A}_n \right)=x_{0}-\triangle\left(\mathit{A}_n\right)$.
   Let $Y^n:=X^n-X^n\textbf{1}_{\mathit{A}_n}$, then $Y^n\in\mathcal{H}_{2}\left(A_n\right)$ and then
   \begin{displaymath}
   \mathbb{E}\left[ \xi Y^n\right]\geq\inf_{Y\in\mathcal{H}_{2}\left(A_n\right)}\mathbb{E}\left[\xi Y \right]
   \end{displaymath}
which means 
\begin{displaymath}
x_0-x_n\geq \triangle\left(A_n\right)=x_0 - x^+\left(A_n\right)
\end{displaymath}
i.e. $x_n\leq x^+\left(A_n\right)$

Let us now focus on the inequality ``$\geq$''. 
Let $A_{n}\in\mathcal{F}$ be such that
\begin{equation}
\nonumber
U\left(\mathit{A}_{n},x_{+}\left(\mathit{A}_{n}\right) \right)\uparrow \sup_{A\in\mathcal{F}}U\left(\mathit{A},x_{+}\left(\mathit{A} \right) \right):=S,\, n\rightarrow +\infty
\end{equation}
By the assumption of the theorem, $x^+(A_n)<\infty$ for all $n$.
Fix $\varepsilon>0$. Our aim is to find, for every $n$, $X_n \in H$ such that 
\begin{align}
U(X_n)\geq U\left(\mathit{A}_{n},x_{+}\left(\mathit{A}_{n}\right) \right) -\varepsilon\label{bonXn}
\end{align} Since $\varepsilon$ is arbitrary it will then follow that $\sup_{X\in H}U(X)\geq S$. 
If $P(A_n)>0$, by Lemma \ref{solvP1} there exists an explicit maximizer of Problem $\Pu$, denoted by $Z(A_n,x^+)$, and 
$U(A_n,x^+) = U(Z(A_n,x^+))$ is continuous in $x^+$. Therefore, we can find $Y_n \in \mathcal H_2(A_n)$ with $\mathbb E[\xi Y_n]$ sufficiently close to $\triangle(A_n)$ so that $U(A_n,x_0-\mathbb E[\xi Y_n])\geq U(A_n,x^+(A_n))-\varepsilon$. Then $X_n:=Z(A_n,x_0-\mathbb E[\xi Y_n]) +Y_n$ satisfies \eqref{bonXn}. If $P(A_n)=0$ then, as we saw in Remark \ref{Remtrivial},  taking $0\in H$ and $X_n=0$ satisfies $U(X_n) = u(0) = U\left(\mathit{A}_{n},x_{+}\left(\mathit{A}_{n}\right) \right)$.

Finally, the fact that $S<\infty$ under Assumption \eqref{ass1a} follows directly from the estimate \eqref{boundU}.

\subsection{Proof of Theorem \ref{corOpt}}
Let $X^*\in\mathit{H}$ be an optimal solution for \eqref{3}, $A^* = \{X^*\geq 0\}$ and $Y^* = X^* \mathbf 1_{X^*<0}$. It is clear that $Y^* \in \mathcal H_2 \left(A^*\right)$. It is also clear that $\mathbb P(A)>0$, since otherwise $\mathbb E[\xi X^*]<x_0$ and one can increase the utility and reduce the risk by increasing $X^*$.  Theorem \ref{10} and the fact that $U(A,x^+)$  is increasing in $x^+$ (Lemma \ref{solvP1}) then give:
\begin{eqnarray}
\nonumber
\sup_{A\in\mathcal F} U \left(A,\xA\right)& = & \sup_{X\in H}\U \\
\nonumber 
& = & U\left(X^*\right) \\
\nonumber
& = & U\left(X^*\textbf{1}_{A^*}\right)\\
\nonumber
& = & U\left(A^*,x_0-\mathbb E[\xi Y^*]\right)\\
\nonumber
& \leq & U\left(A^*,x^+\left(A^*\right)\right)
\end{eqnarray}
which means that $A^*$ achieves the supremum in \eqref{attainsup}. Moreover, since $U(A,x^+)$  is strictly increasing in $x^+$, we get a contradiction unless $x^+(A^*) = x_0-\mathbb E[\xi Y^*]$, which means that $Y^*$ achieves the minimum in $\Pr$.

Conversely, assume that $A^*$ is a maximizer of \eqref{attainsup} and $Y^*$ is a minimizer of $\Pr$. We can then solve Problem $\Pu$ with parameters $(A^*,x_0-\triangle(A^*))$ and we know, by Lemma \ref{solvP1}, that its solution is given by $\left[I\left(\lambda^*\xi \right)^+\right]^+\textbf{1}_{A^*}$. Let then
\begin{displaymath}
X^*:= I\left(\lambda^*\xi \right)\textbf{1}_{A^*} +Y^* 
\end{displaymath}
We have $\rho\left(-(X^*)^-\right)=\rho\left(Y^*\right)\leq\rho_0$ and $\E\left[\xi X^*\right]\leq x_0$, i.e. $X^*\in\mathit{H}$. Using Theorem \ref{corOpt}, we deduce
\begin{eqnarray}
\nonumber
U\left(X^*\right)& = & U\left(X^*\textbf{1}_{A^*}\right) \\
\nonumber
& = & U\left(A^*, x^+\left(A^*\right)\right)\\
\nonumber
& = & \sup_{A\in\mathcal F}U\left(A,\xA\right)\\
\nonumber
& = & \sup_{X\in\mathit{H}}\U.
\nonumber
\end{eqnarray}
 
\subsection{Proof of Theorem \ref{15}}
We will use the methods developed in \cite{jin.zhou.08} (see the proof of Theorem $5.1$). There are however some important differences in our proof which arise in particular due to the presence of risk measures in our context.
 
The theorem will be proved in two steps: in Step 1 we will prove that for every $A\in \mathcal F$, there exists $c \geq 0$ such that $\triangle\left( \mathit{A}\right)\geq\triangle\left(c\right):=\triangle\left(\left\{\xi\leq c\right\}\right)$ so that $x_{+}\left(c\right):=x_0-\triangle\left(\left\{\xi\leq c\right\}\right)\geq x_{+}\left(\mathit{A}\right)$, and in Step 2 we will find, for every $X\in\Au$, an $\hat{X}\in\mathcal{H}_{1}\left(\left\{\xi\leq c\right\},x^+\left(c\right)\right)$ such that $U\left(\hat{X}\right)\geq U\left(X\right)$. We can then conclude that $v\left(c\right):=v\left(\left\{\xi\leq c\right\}\right)\geq v\left(\mathit{A}\right)$

Treating separately the trivial cases as described in Remark \ref{Remtrivial}, we can assume $0<\mathbb{P}\left(\mathit{A} \right)<1$, and set $\alpha=\mathbb{P}(A^c) = 1-\mathbb P(A)$. Let us fix $c\in\left[\underline{\xi},\overline{\xi} \right]$ so that
\begin{displaymath}
\mathbb{P}\left(\xi\leq c  \right)=1-\alpha
\end{displaymath}
This is possible since $\xi$ has no atom. Consider the following sets:
\begin{eqnarray}
\mathit{A}_{1}=\left\{\xi\leq c \right\}\cap\mathit{A}&\mathit{A}_{2}=\left\{\xi>c \right\}\cap\mathit{A}\\
\label{17}
\mathit{B}_{1}=\left\{\xi\leq c \right\}\cap\mathit{A}^c&\mathit{B}_{2}=\left\{\xi>c \right\}\cap\mathit{A}^c\\
\label{18}
\nonumber
\end{eqnarray}
from which it follows $\P\left(\mathit{A}_{2}\right) = \P\left(\mathit{B}_{1}\right)$. If $\mathbb{P}\left(\mathit{A}_{2}\right)=0$ then $\mathit{A}=\left\{\xi\leq c\right\}$, so we can suppose $\mathbb{P}\left(\mathit{A}_{2}\right)>0$.\\
\noindent \textit{Step 1.}\quad
Let $Y \in \mathcal H_2(A)$. Our aim is to  construct $\hat Y \in \mathcal H_2(\{\xi \leq c\})$ with $\mathbb E[\xi \hat Y] = \mathbb E[\xi Y]$ and $\rho(Y) \geq \rho(\hat Y)$. This will imply that $\TA)\geq \triangle(c)$. Introduce the following notation:
\begin{enumerate}
  \item $f_{1}\left(t \right):=\P\left(Y\leq t|\mathit{B}_{1} \right)$
  \item $g_{1}\left(t \right):=\P\left(\xi\leq t |\mathit{A}_{2}\right)$
  \item $Z_{1}=g_{1}\left(\xi\right)$, that is, $\mathcal{L}\left(Z_{1}|\mathit{A}_{2}\right)=\mathcal{U}(\left[0,1 \right])$, because $\xi$ has no atom. 
  \item $W_{1}=f_{1}^{-1}\left(Z_{1} \right)$, that is, the law of $W$ on $A_2$ is the same as the law of $Y$ on $B_1$. 
\end{enumerate}
Let
\begin{equation}
\nonumber
k_{1}:=\left\{\begin{array}{ll}
1 & \textrm{if $W_1=0$ on $\mathit{A}_{2}$}\\
\nonumber
\\[3pt]
\nonumber
\frac{{\mathbb{E}}\left[\xi Y\textbf{1}_{B_{1}}\right]}{{\mathbb{E}}\left[\xi W_1\textbf{1}_{A_{2}} \right]}& \textrm{otherwise}\\
\nonumber
\end{array}\right.
\nonumber
\end{equation}
Observe that since $\xi \leq c$ on $B_1$, and $\xi >c $ on $A_2$, we have that $k\leq 1$. 
Now define
$$
\hat Y = Y \mathbf 1_{B_2} + k W_1 \mathbf 1_{A_2}. 
$$
By definition, $\hat Y = 0$ on $\{\xi \leq c\}$ and $\hat Y \leq 0$ on $\{\xi \leq c\}$. In addition, since $k_1\leq 1$, we easily get that $\mathbb P(-\hat Y>t) \leq \mathbb P(-Y>t)$ for every $t>0$.

\noindent Let $F$ and $\hat F$ be the distribution functions of, respectively, $-Y$ and $-\hat Y$, and $F^{-1}$ and $\hat F^{-1}$ their generalized inverses (defined in \eqref{geninv.eq}). From the above inequality, they satisfy $\hat F^{-1}(u) \leq F^{-1}(u)$ for all $u\in [0,1]$. Let $U$ be a random variable with uniform distribution on $[0,1]$. Since $\rho$ is law invariant,  we obtain that $\rho(\hat Y) = \rho(-\hat F^{-1}(U) ) \leq \rho(-\hat F^{-1}(U) ) =  \rho(Y)\leq \rho_0$ and therefore $\hat Y \in \mathcal H_2(\{\xi \leq c\})$. On the other hand, $\mathbb E[\xi \hat Y] = \mathbb E[\xi Y]$ (this is due to our choice of the constant $k$). Since the choice of $Y$ was arbitrary, this means that $\TA \geq \triangle(c)$. 

\noindent \textit{Step 2.}\quad
Let $X$ be feasible for $\mathcal{P}_{1}$ with parameter $\left(\mathit{A},x_{+}\left(\mathit{A}\right)\right)$, and define
\begin{enumerate}
  \item $f_{2}\left(t \right):=\mathbb{P}\left(X\leq t|\mathit{A}_{2} \right)$
  \item $g_{2}\left(t \right):=\mathbb{P}\left(\xi\leq t |\mathit{B}_{1}\right)$
  \item $Z_{2}=g_{2}\left(\xi\right)$
  \item $W_{2}=f_{2}^{-1}\left(Z_{2} \right)$, that is, the law of $W_2$ on $B_1$ is the same as the law of $X$ on $A_2$. 
\end{enumerate}
Let
\begin{displaymath}
k_{2}:=\left\{\begin{array}{cc}
1 & \textrm{if $W_{2}=0$ on $\mathit{B}_{1}$}\\
\frac{\mathbb{E}\left[\xi X\textbf{1}_{\mathit{A}_{2}} \right]}{\mathbb{E}\left[\xi W_{2}\textbf{1}_{\mathit{B}_{1}} \right]}& \textrm{otherwise}
\end{array}\right.
\end{displaymath}
Note that now, $k_2\geq 1$. We define a new random variable $\hat X$ by
\begin{align}
\hat{X}:=X\textbf{1}_{\mathit{A}_{1}}+k_{2}W_{2}\textbf{1}_{\mathit{B}_{1}} + \frac{x_{+}\left(c\right)-\xA}{\E\left[\xi\textbf{1}_{\left\{\xi\leq c\right\}} \right]}\textbf{1}_{\left\{\xi\leq c\right\}}\\
\nonumber
\end{align}
We have $\E\left[\xi \hat{X} \right]=x^{+}\left(c\right)$ and it is easy to see that $\hat{X}\in\mathcal{H}_1 \left(\left\{\xi\leq c\right\},x^+\left(c\right)\right)$.
Moreover, since $k_2\geq 1$, a simple computation shows that
\begin{displaymath}
\P\left(\hat{W}>t\right)\geq \P\left(X>t\right)
\end{displaymath}
By definition,
\begin{align}
&U\left( X\right)=\E\left[u\left(X^{+}\right) \right]=\int^{+\infty}_{0}\mathbb{P}\left(X^{+}>u^{-1}\left(t\right) \right)dt\\
\nonumber
&U\left( \hat{W}\right)=\E\left[u\left( \hat{W}^{+}\right) \right]=\int^{+\infty}_{0}\P\left(\hat{W}^{+}>u^{-1}\left(t\right) \right)dt
\end{align}
but since $u^{-1}\left(t\right)$ is positive,
\begin{align}
\nonumber
\left\{ \hat{W} ^{+}>u^{-1}\left(t\right)\right \}=\left\{ \hat{W} >u^{-1}\left(t\right) \right\}\\
\nonumber
\left\{X^{+}>u^{-1}\left(t\right)\right\}=\left\{X >u^{-1}\left(t\right)\right\},
\nonumber
\end{align}
which enables us to conclude that $U\left(X\right) \leq U\left( \hat{X}\right)$. 

%%%%%%%%%%%%%%%%%%%%%%%%%%%%%%%%%%%%%%%%%%%%%%%%%%%%%%%%%%%%%%%%%%%%%%%%%%%%%%%%%%%%%%%%%%%%%%%%%%%%%%%%%%%%%

\subsection{Proof of Theorem \ref{entropic.exist}}

%%%%%%%%%%%%%%%%%%%%%%%%%%%%%%%%%%%%%%%%%%%%%%%%%%%%%%%%%%%%%%%%%%%%%%%%%%%%%%%%%%%%%%%%%%%%%%%%%%%%%%%%%%%%%
\begin{proof}
The proof is just a simple application of Theorems \ref{corOpt}, \ref{15},  Lemma \ref{solvP1} and Lagrangian methods.

Fix $c$ and consider the problem:
\begin{displaymath}
\left\{\begin{array}{ll}
\inf\mathbb{E}\left[\xi Y\right]&\textrm{s.t.}\\
\rho\left(Y\right)\leq\rho_{0}&\\
Y=0\quad \textrm{on $\mathit{A}$},\quad Y\leq 0 \quad\textrm{on $\mathit{A}^{c}$}&\textrm{and} \\
\end{array}\right.
\end{displaymath}
where $\mathit{A}=\left\{\xi\leq c\right\}$\\
Working on the new space $\left({A}^{c},\hat{\mathcal{F}}:= \{B\cap A^c, B \in \mathcal F\},\hat{\mathbb{P}}:=\mathbb P(\cdot| A^c) \right)$, we can transform this minimization into
\begin{displaymath}
\left\{\begin{array}{cc}
\inf\alpha\left(c\right)\hat{\mathbb{E}}\left[\xi W\right] &\textrm{s.t.}\\
\\[4pt]
\hat{\mathbb{E}}\left[ \exp\left(-\frac{W}{\beta}\right)\right]\leq\delta\left(c\right),& W\leq 0\\
\end{array}\right.
\end{displaymath}
where 
\begin{equation}
\nonumber
\delta\left(c\right) =\frac{e^{\frac{\rho_{0}}{\beta}}+\alpha\left(c\right)-1}{\alpha\left(c\right)}.
\end{equation}
Using Lagrangian methods we can find the unique optimal solution:
\begin{equation}
\nonumber
W^{*}\left(c\right):=-\beta\left[\log\left(\frac{\beta}{\eta\left(c\right)}\xi\right)\right]^+
\end{equation}
where $\eta\left(c\right)$ is the unique solution of:
$$\mathbb E\left[\left(\frac{\beta \xi}{\eta(c)}\vee 1\right) \mathbf 1_{\xi > c}\right] = e^{\frac{\rho_0}{\beta}} + \alpha(c)-1,$$
and so
\begin{equation}
\nonumber
Y^{*}\left(c\right):=W^*\left(c\right)\textbf{1}_{\left\{\xi>c\right\}}.
\end{equation}
A simple calculation then gives:
$$\triangle\left(c\right)= -\beta \mathbb E\left[\xi \log\left(\frac{\beta \xi }{\eta(c)}\vee 1\right)\right].$$
If now we set $x_{+}\left(c\right)=x_{0}-\triangle\left(c\right)$, by Lemma \ref{solvP1} , Problem $\Pu$ with parameters $\left(\left\{\xi\leq c\right\},x^+\left(c\right)\right)$ can be solved and its unique solution is
\begin{equation}
\nonumber
X\left(c\right)=I\left(\lambda\left(c\right) \xi \right) \textbf{1}_{\left\{\xi\leq c\right\}},
\end{equation}
where, by \eqref{Lag}, 
\begin{displaymath}
\E\left[\xi I\left(\lambda\left(c\right)\xi\right) \textbf{1}_{\left\{\xi\leq c\right\}} \right]=x^+\left(c\right).
\end{displaymath}
Using Theorem \ref{15}, the optimal $c^*$ is the maximizer of the function
\begin{displaymath}
c\rightarrow \E\left[u\left(I\left(\lambda\left(c\right)\xi\right)  \right)\textbf{1}_{\left\{\xi\leq c\right\}} \right].
\end{displaymath} 
\end{proof}

%%%%%%%%%%%%%%%%%%%%%%%%%%%%%%%%%%%%%%%%%%%%%%%%%%%%%%%%%%%%%%%%%%%%%%%%%%%%%%%%%%%%%%%%%%%%%%%%%%%%%%%%%%%%%

\subsection{Proof of Lemma \ref{explicit_TA}}

%%%%%%%%%%%%%%%%%%%%%%%%%%%%%%%%%%%%%%%%%%%%%%%%%%%%%%%%%%%%%%%%%%%%%%%%%%%%%%%%%%%%%%%%%%%%%%%%%%%%%%%%%%%%%
\begin{proof}
In order to compute $\TA$ we reformulate Problem $\Pr$ in terms of the conditional distribution function of $Y\in\mathcal H_2\left(A\right)$ on $A^c$. Introduce a new probability $\hat {\mathbb P}$ via $\frac{d \hat {\mathbb P}}{ d {\mathbb P}} = \frac{\mathbf 1_{A^c}}{\alpha_A}$. Let $\hat F_Y$ be the distribution function of $Y$ under this probability and $\hat F_Y^{-1}$ its generalized inverse. Using this new probability we can rewrite the ingredients of our problem as 
$$
\mathbb E[\xi Y] = \alpha_A \hat{\mathbb E}[\xi Y]
$$
and
\begin{align*}
CVAR_\beta(Y) &= - \frac{1}{\beta} \int_0^\beta F^{-1}_Y (u) = -\frac{1}{\beta} \int_0^{\beta \wedge \alpha_A} \hat F^{-1}_Y (u/\alpha_A) du\\  &= - \frac{\alpha_A}{\beta} \int_0^{\frac{\beta}{\alpha_A}\wedge 1} \hat F^{-1}_Y(u) du. 
\end{align*}
and then using \eqref{rho_mu} we obtain
\begin{align*}
\rho_\mu(Y) &= -\alpha_A \int_0^1 \phi\left(\alpha_A u\right)\hat F^{-1}_Y(u) du. 
\end{align*}
To express $\hat{\mathbb E}[\xi Y]$, we use the following well known result:
\begin{lemma}
\label{copulae}
Let $F_1$ and $F_2$ be distribution functions on $[0,\infty)$. Then
\begin{align*}
\sup_{X\sim F_1,\ Y\sim F_2} \mathbb E[XY] = \int_{0}^1 F_1^{-1}(u) F_2^{-1}(u) du.
\end{align*}
\end{lemma}
Using this lemma, Problem $\Pr$ can be expressed as
\begin{align}
&\TA =\alpha_A \inf \int_0^1 \hat F^{-1}_\xi (u) \hat F^{-1}_Y(1-u)du \label{p2mod1} \\
&\text{subject to} \quad - \alpha_A\int_0^1 \phi\left(\alpha_A u\right)\hat F^{-1}_Y(u) du \leq \rho_0,\label{p2mod2}
\end{align}
where the $\inf$ is taken over all generalized inverse distribution functions $\hat F^{-1}_Y$ of non-positive random variables. Such a function can always be written as
\begin{align}
\hat F^{-1}_Y\left(u\right):=-\int_u^1 \zeta\left(du\right),
\label{F_inv}
\end{align}
where $\zeta$ is a positive measure on $[0,1]$. Using Fubini's theorem, we can rewrite problem \eqref{p2mod1}--\eqref{p2mod2} in terms of this measure:
\begin{align*}
&\TA =-\alpha_A \sup \left(\int_0^1 \zeta(ds) \int_{0}^s\hat F^{-1}_\xi (1-u)du\right) \\
&\text{subject to} \quad  \alpha_A \left( \int_0^1 \zeta(ds) \int_0^s\phi\left(\alpha_A u\right)du \right)\leq \rho_0.
\end{align*}
The solution of this problem can easily be shown to be a point mass: $\zeta = h \delta_x$ where $h\geq 0$ and $x\in[0,1]$ can be found from
\begin{align}
&\TA =- \alpha_A \sup \left(h\int_0^x\hat F^{-1}_\xi (1-u)du \right) \label{p2bismod1} \\
&\text{subject to} \quad \alpha_A h\int_0^x \phi\left(\alpha_A u\right)du = \rho_0, \label{p2bismod2}
\end{align}
The constraint \eqref{p2bismod2} gives us 
\begin{align*}
h=h\left(x\right)=\frac{\rho_0}{\alpha_A \int_0^x \phi\left(\alpha_A s\right)ds}
\end{align*}
and using definition \eqref{rx} we get
\begin{eqnarray*}
\TA &=&-\alpha_A \sup_{x\in [0,1]} \left(\frac{\rho_0}{\alpha_A \int_0^x \phi\left(\alpha_A s\right)ds}\int_0^x\hat F^{-1}_\xi (1-u)du \right) \\
\\[3pt]
& =& -\rho_0 \sup_{x\in\left[0,1\right]}r\left(x\right) \\
\end{eqnarray*}
The function $r$ is differentiable on $(0,1]$ and may only have a singularity at $x=0$; using l'H\^opital's rule, we get
\begin{align*}
r\left(0^+\right)=\lim_{x\rightarrow 0}\frac{\hat F^{-1}_\xi (1-x)}{\phi\left(x\right)}
\end{align*}
So $\TA<+\infty$ if and only if $r$ is bounded on $\left[0,1\right]$, which is true if and only if $r\left(0^+\right)<+\infty$.
\end{proof}

%%%%%%%%%%%%%%%%%%%%%%%%%%%%%%%%%%%%%%%%%%%%%%%%%%%%%%%%%%%%%%%%%%%%%%%%%%%%%%%%%%%%%%%%%%%%%%%%%%%%%%%%%%%%%

\subsection{Proof of Corollary \ref{max_c_spectr}}

%%%%%%%%%%%%%%%%%%%%%%%%%%%%%%%%%%%%%%%%%%%%%%%%%%%%%%%%%%%%%%%%%%%%%%%%%%%%%%%%%%%%%%%%%%%%%%%%%%%%%%%%%%%%%

\begin{proof}
In order to make the dependence on $c$ explicit, we introduce the notation
\begin{align*}
&\triangle\left(c\right):=-\rho_0 \max_{x\in\left[0,1\right]} R\left(x,c\right)
\end{align*}
where
\begin{align*}
&R\left(x,c\right):=\frac{\alpha\left(c\right)\int_0^x \hat F^{-1}_\xi\left(1-u\right)du}{\int_0^{\alpha\left(c\right)x}\phi\left(u\right)du}
\end{align*}
Noting that $\hat F_\xi^{-1} (1-u) = F^{-1}_\xi(1-\alpha(c) u)\geq c$ and making a change of variable, 
\begin{align*}
R\left(x,c\right)&=\frac{\E\left[\xi \textbf{1}_{\left\{c<\xi \right\}}\textbf{1}_{\left\{\hat F^{-1}_\xi\left(1-x\right)<\xi \right\}} \right]}{\int_0^{\alpha\left(c\right)x}\phi\left(u\right)du} = \frac{\E\left[\xi \textbf{1}_{\left\{c<\xi \right\}}\textbf{1}_{\left\{\hat F^{-1}_\xi\left(1-x\right)<\xi \right\}} \right]}{\int_0^{\alpha\left(c\right)x}\phi\left(u\right)du}\\
& = \frac{\E\left[\xi \textbf{1}_{\left\{ F^{-1}_\xi\left(1-\alpha(c)x\right)<\xi \right\}} \right]}{\int_0^{\alpha\left(c\right)x}\phi\left(u\right)du} = \frac{\E\left[\xi \textbf{1}_{\left\{ 1-F_\xi(\xi)<\alpha(c) x \right\}} \right]}{\int_0^{\alpha\left(c\right)x}\phi\left(u\right)du}
\end{align*}
The function $\Delta (c)$ can then be rewritten as
$$
\Delta(c) = -\rho_0 \max_{0\leq z \leq \alpha(c)} R(z),\quad R(z):=\frac{\E\left[\xi \textbf{1}_{\left\{ 1-F_\xi(\xi)<z \right\}} \right]}{\int_0^{z}\phi\left(u\right)du}
$$
\end{proof}

%%%%%%%%%%%%%%%%%%%%%%%%%%%%%%%%%%%%%%%%%%%%%%%%%%%%%%%%%%%%%%%%%%%%%%%%%%%%%%%%%%%%%%%%%%%%%%%%%%%%%%%%%%%%%

\subsection{Proof of Theorem \ref{cond_spect}}

%%%%%%%%%%%%%%%%%%%%%%%%%%%%%%%%%%%%%%%%%%%%%%%%%%%%%%%%%%%%%%%%%%%%%%%%%%%%%%%%%%%%%%%%%%%%%%%%%%%%%%%%%%%%%

\begin{proof}
From Theorem \ref{15} we need to maximize the function $c\mapsto v(c)$ over $c\in [\underline \xi,\, \overline\xi ]$. Assume that $v(c)$ achieves its maximum at the point $c^*$ such that $\Delta(c^*) = -\rho R(z)$ with $z< \alpha(c)$ and let $c' = \alpha^{-1}(z)$. Then, $\Delta(c)$ is constant on the interval $[c,c']$, which means that  $x^+(c) = x^+(c')$,
$$
\mathcal{H}_1\left(\left\{\xi\leq c\right\},x^+(c)\right)\subset\mathcal{H}_1\left(\left\{\xi\leq c'\right\},x^+(c')\right)
$$
and therefore $v(c)\leq v(c')$. This argument shows that the solution of the optimization problem appearing in the right-hand side of \eqref{16} does not change if we replace the expression for $\Delta(c)$ given by Corollary \ref{max_c_spectr} by 
$$
-\rho_0 R(\alpha(c)) = -\frac{\rho_0\E[\xi \mathbf 1_{\xi>c}]}{\int_0^{\P[\xi>c]} \phi(u)du}.
$$
Applying Lemma \ref{solvP1} we then find 
$$
v(c) = \E[u(I(\lambda(c)\xi))\mathbf{1}_{\xi \leq c}],
$$
where
$$
\E[\xi I(\lambda(c)\xi) \mathbf{1}_{\xi \leq c} ] = x_0 + \frac{\rho_0\E[\xi \mathbf 1_{\xi>c}]}{\int_0^{\P[\xi>c]} \phi(u)du}.
$$
If there exists a $c^*$ with $\P(\xi>c^*)>0$ which maximizes the value function $c\to v(c)$ then the optimal contingent claim is given by
$$
X^* =I(\lambda(c^*)\xi) \mathbf{1}_{\xi \leq c^*}- \frac{\rho_0}{\int_0^{\P[\xi > c^*]} \phi(u)du}\mathbf 1_{\xi>c^*}.
$$
where 
$$
- \frac{\rho_0}{\int_0^{\P[\xi > c^*]} \phi(u)du}\mathbf 1_{\xi>c^*}.
$$ 
is the optimal solution of Problem $\Pr$ corresponding to $\left\{\xi\leq c^* \right\}$, which can be deduced from the proof of Lemma \ref{explicit_TA}.
\end{proof}

\section*{Acknowledgement}
This research is part of the Chair {\it Financial Risks} of the {\it Risk
Foundation} sponsored by Soci\'et\'e G\'en\'erale, the Chair {\it
Derivatives of the Future} sponsored by the {F\'ed\'eration Bancaire
Fran\c{c}aise}, and the Chair {\it Finance and Sustainable Development}
sponsored by EDF and Calyon.

%\bibliographystyle{chicago}
%\bibliography{portfolio_insurance}

\end{document}